\documentclass{ws-procs9x6}

\usepackage{psfrag}

\psfrag{QF2F1}{{\footnotesize $\sqrt{Q^2}F_2^p/F_1^p$}}

\psfrag{Q2}{{\footnotesize $Q^2$}}

\psfrag{GMDELTA}{{\footnotesize $G^*_M/(3 G_D)$}}

\psfrag{REM}{{\footnotesize $R_{EM} = -G^*_E/G^*_M$}}

\psfrag{RSM}{{\footnotesize $R_{SM} \propto  -G^*_C/G^*_M$}}

\psfrag{F1p}{{\footnotesize $F_1^p/G_D$}}

\psfrag{F2p}{{\footnotesize $F_2^p/(\mu_p G_D)$}}

\newcommand{\xx}{x}

\newcommand{\D}{{\cal D}}

\newcommand{\up}{{\uparrow}}
\newcommand{\down}{{\downarrow}}

\newcommand{\ep}{\varepsilon}
\newcommand{\ga}{\gamma}

\newcommand{\bra}[1]{\left\langle #1 \right|}
\newcommand{\ket}[1]{\left| #1 \right\rangle}

\begin{document}

\title{NUCLEONS ON THE LIGHT CONE:\\ THEORY AND PHENOMENOLOGY OF BARYON DISTRIBUTION AMPLITUDES}

\author{V. M. BRAUN$^*$}

\address{Institut f{{\"u}}r Theoretische Physik, \\
 Universit{{\"a}}t  Regensburg,\\
 D-93040 Regensburg, Germany\\
$^*$E-mail: vladimir.braun@physik.uni-regensburg.de}

\begin{abstract}
This is a short review of the theory and phenomenology of baryon distribution amplitudes,
including recent applications to the studies of nucleon form factors at intermediate 
momentum transfers using the light-cone sum rule approach.
\end{abstract}


\bodymatter

\section{Introduction}\label{sec1}
In the next generation of experiments in hadron physics there is a tendency to
go for more and more exclusive  channels. 
One main reason for this is that one has understood only in
recent years how much one can learn from reactions like 
deeply virtual Compton scattering (DVCS) about the
internal hadron structure and especially the spin structure. 
All future plans also call for very high luminosity and would therefore be perfectly suited for 
the investigation of exclusive and semi-exclusive reactions with and without 
polarization. 

Main question which has to be addressed at this stage is whether 
studies of hard exclusive processes can be made fully quantitative.
The classical theoretical 
framework for the calculation of hard exclusive processes in QCD
is based on QCD factorization \cite{Chernyak:1977as,Radyushkin:1977gp,Lepage:1979zb}. This
approach introduces a concept of hadron distribution amplitudes (DAs)
as fundamental nonperturbative 
functions describing the hadron structure in rare parton configurations
with a fixed number of Fock constituents at small transverse 
separation.  DAs are ordered by increasing twist. For example, 
the leading-twist-2 meson DA $\phi_{2;P}$ describes the momentum
 distribution of the valence quarks in the meson $P$ and is related to the
 meson's Bethe--Salpeter wave function $\phi_{P,BS}$ by an integral
 over transverse momenta:
$
\phi_{2;P}(\xx,\mu) = Z_2(\mu) \int^{|k_\perp| < \mu} \!\!d^2 k_\perp\,
\phi_{P,BS}(\xx,k_\perp).
$
Here $\xx$ is the quark momentum fraction, $Z_2$ is the renormalization factor (in the light-cone gauge) 
for the quark-field operators in the wave function, and $\mu$ denotes the renormalization scale.
Higher-twist DAs are much more numerous and describe either
contributions of ``bad'' components in the wave function, or contributions
of transverse motion of quarks (antiquarks) in the leading-twist
components, or contributions of higher Fock states with additional
gluons and/or quark--antiquark pairs. Within the hard-rescattering picture, 
the corresponding contributions to the hard exclusive reactions are
suppressed by a power (or powers) of the large   
momentum $Q$ and usually have received less attention.

The distribution amplitudes are equally important and 
to a large extent complementary to conventional 
parton distributions which correspond to one-particle probability
distributions for the parton momentum fraction in an 
{\em average}  configuration. They are, however, much 
less studied: The direct experimental information 
is only available for the pion distribution amplitude and comes
from the CLEO measurement \cite{Gronberg:1997fj} of the $\gamma^*\gamma\pi$
transition form factor. 
In this talk I present a short review of the theoretical status 
of baryon DAs, mainly those of the nucleon. I describe the basic
theoretical framework and discuss how the nucleon DAs can be related 
to the experimental measurements of form factors at accessible momentum 
transfers within the light-cone sum rule (LCSR) framework.

\section{General framework}

\subsection{Definitions}

Nucleon DAs are most conveniently defined as nucleon-to-vacuum transition matrix
elements of nonlocal light-ray three-quark operators. In order to facilitate the 
power counting it is usually convenient to choose a light-like vector $z_\mu$ orthogonal to the 
large momentum $q_\mu$ involved in the problem: 
      $ q\cdot z =0\,,\qquad z^2 =0 $\,.
The nucleon momentum $P_\mu, P^2 =m_N^2$ can be used to introduce the second light-like vector 
$p_\mu = P_\mu  - \frac{1}{2} \, z_\mu \frac{m_N^2}{P\cdot z}\,,$ 
$p^2=0$,
so that $P \to p$ if the nucleon mass can be neglected, $m_N \to 0$.

The nucleon leading-twist-three DA $\Phi_N(\xx_i)$ can be defined as
\cite{Chernyak:1984bm,Chernyak:1983ej,Braun:1999te,Braun:2000kw}
\begin{eqnarray}
\lefteqn{\hspace*{-1cm}\langle 0| \ep^{ijk} 
\left(u^{\up}_i(a_1 z) C \!\not\!{z} u^{\down}_j(a_2 z)\right)  
\!\not\!{z} 
d^{\up}_k(a_3 z) |N(P)\rangle
=}
\nonumber\\
&&{}\hspace*{2.5cm}=\,- \frac12 pz\! \not\!{z} N^\up(P) \int\! \D\xx 
e^{-i pz \sum \xx_i a_i} {\Phi_N(\xx_i)}
\label{phi3}
\end{eqnarray}
where 
$  \int \D\xx = \int_0^1 d\xx_1\, d\xx_2\, d\xx_3\,\delta\left(1-\xx_1-\xx_2-\xx_3\right)$,
$\xx_i$ correspond to quark momentum fractions, $C$ is the charge-conjugation matrix,  
$N(P)$ is the Dirac spinor and the arrows correspond to the helicity projections
   $ q^{\up(\down)} = ({1\pm\ga_5})/{2}\,q$. 

The definition in (\ref{phi3}) is equivalent to the following representation for the 
three-quark component of the proton wave function \cite{Lepage:1979zb}
\begin{equation}
 |p\uparrow\rangle =  \int \frac{  \D\xx\, \Phi_N(x_i)}{2\sqrt{24 x_1 x_2 x_3}} 
 \left\{ |u^\up(x_1)u^\down(x_2)d^\up(x_3)\rangle -  |u^\up(x_1)d^\down(x_2)u^\up(x_3)\rangle
 \right\},  
\end{equation}
where the standard relativistic normalization of spinors is implied. 
One often writes 
$
 \Phi_N(1,2,3) = V_1(1,2,3) - A_1(1,2,3)
$,
where $V_1(\xx_1,\xx_2,\xx_3) = V_1(\xx_2,\xx_1,\xx_3,)$ and $A_1(\xx_1,\xx_2,\xx_3) = -A_1(\xx_2,\xx_1,\xx_3,)$ 
correspond to the symmetric and the antisymmetric part of $\Phi_N(\xx_i)$ w.r.t. the interchange of the 
u-quark momenta, respectively.  
The definitions of the leading-twist DAs of other baryons of the octet 
can be found in Ref.~\refcite{Chernyak:1987nv}.

{}For higher twists,
there exists an important conceptual difference between mesons and baryons.
For mesons, all effects of the transverse motion of quarks in the valence quark-antiquark state can 
be rewritten in terms of higher Fock state contributions by using QCD equations of motion (EOM).
Since quark-antiquark-gluon admixture in meson wave functions turns out to be numerically small, 
the transverse momentum contributions are small as well, and the higher-twist contributions 
to hard exclusive reactions involving mesons are dominated in most cases by meson mass corrections, 
see e.g. Ref.~\refcite{Ball:1998ff}. For baryons, EOM are not sufficient to eliminate higher-twist 
three-quark DAs in favor of the components with extra gluons, so that the former present genuine new degrees 
of freedom. A systematic classification of such contributions is carried out in Ref.~\refcite{Braun:2000kw}.
One finds that to the twist-four accuracy there exist three independent DAs: 
\begin{eqnarray}
\lefteqn{\hspace*{-1cm}\langle 0| \ep^{ijk} 
\left(u^{\up}_i(a_1 z) C \!\not\!{z} u^{\down}_j(a_2 z)\right)  
\!\not\!{p} 
d^{\up}_k(a_3 z) |N(P)\rangle = }
\nonumber\\
 &&{}\hspace*{2.5cm}=\,- \frac12 pz \!\not\!{p} N^\up(P) \int\!\! \D\xx\, 
e^{-i pz \sum \xx_i a_i}
{\Phi_4(\xx_i)}\,,
\nonumber\\
\lefteqn{\hspace*{-1cm}\langle{0}| \ep^{ijk} 
\left(u^{\up}_i(a_1 z) C\!\not\!{z} \gamma_\perp \!\!\not\!{p}\, u^{\down}_j(a_2 z)\right)  
\gamma_\perp\!\not\!{z} 
d^{\down}_k(a_3 z) |{N(P)}\rangle =}
\nonumber\\ 
&&{}\hspace*{2.5cm}=\, -m_N\,pz\not\!{z} N^\up(P) \int\! \D\xx\, 
e^{-i pz \sum \xx_i a_i}
{\Psi_4(\xx_i)}\,,
\nonumber\\
\lefteqn{\hspace*{-1cm}\bra{0} \ep^{ijk} 
\left(u^{\up}_i(a_1 z) C \!\not\!{p}\!\not\!{z}  u^{\up}_j(a_2 z)\right)  
\!\not\!{z} 
d^{\up}_k(a_3 z) \ket{N(P)} =}
\nonumber\\
&& {}\hspace*{2.5cm}=\, \phantom{-}\frac12 m_N  pz \!\not\!{z} N^\up(P) \int\!\! \D\xx\, 
e^{-i pz \sum \xx_i a_i}
{\Xi_4(\xx_i)}\,.
\label{twist4}
\end{eqnarray}
In addition, there exist three twist-5 and one twist-6 three-quark DA, 
which do not involve new parameters to this accuracy,  and can be expressed 
in terms of twist-3,4 DAs.

Note that in the approach of Ref.~\refcite{Braun:2000kw} the higher-twist DAs
are introduced as matrix elements of light-ray operators involving ``minus''
components of the quark field operators. All transverse 
degrees of freedom are eliminated. There exists an alternative approach
\cite{Ji:2002xn} in which only ``plus'' components are involved, but the
transverse momentum dependence is retained. Both techniques are probably 
equivalent but the precise connection has not been worked out yet.

\subsection{Conformal expansion}

A convenient tool to study DAs is provided by conformal expansion 
\cite{Brodsky:1980ny,Makeenko:1980bh,Ohrndorf:1981qv,Braun:1989iv,Mueller:1993hg,Belitsky:1998gc,Braun:2003rp}.
The underlying idea is similar to partial-wave decomposition in quantum mechanics and allows one to separate
transverse and longitudinal variables in the Bethe--Salpeter wave function.  The
dependence on transverse coordinates is traded for the scale dependence
of the relevant operators and is governed by
renormalization-group equations, the dependence on the longitudinal
momentum fractions is described in terms of irreducible
representations of the corresponding symmetry group, the collinear
conformal group SL(2,$\mathbb R$). The conformal partial-wave expansion is
explicitly consistent with the equations of motion since the latter are
not renormalized. It thus makes maximum use of the symmetry
of the theory to simplify the dynamics. 

To construct the conformal expansion for an arbitrary multiparticle
distribution, one first has to decompose each constituent field into
components with fixed Lorentz-spin projection onto the
light-cone. Each such component has conformal spin
$
j=\frac{1}{2}\, (l+s),
$
where $l$ is the canonical dimension  and $s$ the (Lorentz-) spin
projection. In particular, $l=3/2$ for quarks and $l=2$ for gluons.
The  quark field is decomposed as $\psi_+ \equiv
\Lambda_+\psi$ and $\psi_-=
\Lambda_-\psi$ with spin projection operators $\Lambda_+ = \frac{\not{p}\,\not{z}}{2pz}$ and 
 $\Lambda_- = \frac{\not{z}\,\not{p}}{2pz}$, corresponding to
$s=+1/2$ and $s=-1/2$, respectively.
Note that the ``minus'' components of quark fields that contribute to higher-twist DAs
correspond to the negative spin projection and thus lower conformal spin. 
The three-particle states built of quark with definite Lorentz-spin
projection can be expanded in irreducible  representations of SL(2,$\mathbb R$) 
with increasing conformal spin.
The explicit expression for the DA with the lowest possible conformal spin
 $j=j_1+j_2+j_3$, the so-called asymptotic DA, is
\begin{equation}
\phi_{as}(\xx_1,\xx_2,\xx_3) =
\frac{\Gamma(2j_1+2j_2 +2j_3)}{\Gamma(2j_1)\Gamma(2j_2) \Gamma(2j_3)}\,
\xx_1^{2j_1-1}\xx_2^{2j_2-1}\xx_3^{2j_3-1}.
\label{eq:asymptotic}
\end{equation}
{}For the leading twist DAs $j_1=j_2=j_3=1$ reproducing the familiar result
\begin{equation}
  \Phi^{\rm as}_N(\xx_1,\xx_2,\xx_3) = 120 \xx_1\xx_2\xx_3\,.
\end{equation}
The nucleon DA can be expanded in the sum over irreducible representations with higher spin 
$N+3$. For example for leading twist
\begin{eqnarray}
 \Phi_N(\xx_i) &=&  \Phi^{\rm as}_N(\xx_i)\sum_{N=0}^\infty \sum_{n=0}^N \varphi_{N,n}(\mu)
  \Psi^{(12)3}_{N,n}(\xx_i)
\label{expand}
\end{eqnarray}
 where \cite{Braun:1999te} 
\begin{eqnarray}
  \Psi^{(12)3}_{N,n}(\xx_i) = (\xx_1+\xx_2)^n P^{(2n+3,1)}_{N-n} (\xx_3-\xx_1-\xx_2)
    C^{3/2}_n\left(\frac{\xx_1-\xx_2}{\xx_1+\xx_2}\right)
\label{basis}
\end{eqnarray}
where the constraint
$\sum_{k=1}^3 \xx_k=1$ is implied and
$C_n^{3/2}(x)$ and $P^{\alpha,\beta}_n(x)$ are Gegenbauer and Jacoby polynomials,
respectively. 
The superscript $(12)3$ stands for the order in which the conformal
spins of the three quarks are summed to form the total spin $N+3$: First the 
$u$-quark spins are summed to the total spin $n+2$, and then the $d$-quark spin is added.
This order is of course arbitrary; the functions 
$\Psi^{(12)3}_{N,n}$ and e.g. $\Psi^{1(23)}_{N,n}$ are related with each other through
the Racah $6j$ symbols of the $SL(2)$ group, see Ref.~\refcite{Braun:1999te}   
and Appendix B in Ref.~\refcite{Braun:2001qx} for details.
The basis functions in (\ref{basis}) are mutually orthogonal w.r.t. the conformal 
scalar product (\ref{eq:asymptotic}) and are more convenient than Appell polynomials 
used in earlier studies \cite{Peskin:1979mn,Brodsky:1980ny,Ohrndorf:1981qv,Tesima:1981ud,Braun:1989iv,Nyeo:1992qd,Stefanis}.

The explicit expression for the conformal expansion of the leading-twist proton DA
(\ref{phi3}) to the next-to-leading conformal spin accuracy ($N=0,1$) reads
\begin{eqnarray}
\Phi_N(x_i,\mu) &=& 120 x_1 x_2 x_3 \left[\phi_3^0 + (x_1 - x_2) \phi_3^-
+\phi_3^+ (1- 3 x_3)\right]\,,
\label{DA3}
\end{eqnarray}
and the twist-4 DAs (\ref{twist4}) to the same accuracy are given by
\begin{eqnarray} 
\Phi_4(x_i) &=& 24 x_1 x_2 
\left[\phi_4^0 + \phi_4^- (x_1 - x_2) + 
\phi_4^+ (1- 5 x_3)\right] \,,
\nonumber\\
\Psi_4(x_i) &=& 24 x_1 x_3 
\left[\psi_4^0 + \psi_4^- (x_1 - x_3) + 
\psi_4^+ (1- 5 x_2)\right]\,,
\nonumber\\
\Xi_4(x_i) &=& 24 x_2 x_3 
\left[\xi_4^0  + \xi_4^-  (x_2 - x_3) + 
\xi_4^+(1- 5 x_1)\right] \,,
\label{DA4}
\end{eqnarray}
The twelve coefficients  $\phi_3^0\ldots \xi^+_4$
can be expressed in terms of eight independent
non-perturbative parameters $f_N, \lambda_1, \lambda_2, f_1^u, f_1^d, f_2^d, A_1^u, V_1^d$
corresponding to matrix elements of local operators. 
One obtains \cite{Braun:2000kw}
\begin{eqnarray}
\phi_3^0 &=& f_N \,, \quad  
  \phi_3^- = \frac{21}{2}f_N A_1^u\,,\quad  \phi_3^+ = \frac72(1-3V_1^d)\,  
\label{par3}
\end{eqnarray}
for the leading twist, and
\begin{eqnarray}
\phi_4^0 &=& \frac{1}{2} \left(\lambda_1 + f_N\right) \,,\quad 
\xi_4^0  = \frac{1}{6} \lambda_2\,,\quad \psi_4^0 = \frac{1}{2} \left(f_N-\lambda_1\right)  
\nonumber\\
\phi_4^- &=& \frac{5}{4} \left(\lambda_1(1- 2 f_1^d -4 f_1^u) 
+ f_N( 2 A_1^u - 1)\right) \,,
\nonumber \\
\phi_4^+ &=& \frac{1}{4} \left( \lambda_1(3- 10 f_1^d) 
- f_N( 10  V_1^d - 3)\right)\,,
\nonumber \\
\psi_4^- &=& - \frac{5}{4} \left(\lambda_1(2- 7 f_1^d + f_1^u) 
+ f_N(A_1^u + 3 V_1^d - 2)\right) \,,
\nonumber \\
\psi_4^+ &=& - \frac{1}{4} \left(\lambda_1 (- 2 + 5 f_1^d + 5 f_1^u) 
+ f_N( 2 + 5 A_1^u - 5 V_1^d)\right)\,,
\nonumber \\
\xi_4^- &=& \frac{5}{16} \lambda_2(4- 15 f_2^d)\,,\qquad
\xi_4^+ = \frac{1}{16} \lambda_2 (4- 15 f_2^d)\,.
\label{par4}
\end{eqnarray}
for the twist-four DAs, respectively. 
Note that the truncation of the conformal expansion at first order tacitly implies 
an assumption that this expansion is well convergent at least as a distribution 
in mathematical sense: after convolution with a smooth test function. 

\subsection{Scale dependence and Complete Integrability}

The scale dependence of the nonperturbative coefficients $\varphi_{N,n}(\mu)$  in (\ref{expand})
is obtained by 
the diagonalization of the mixing matrix for the three-quark operators
\begin{eqnarray}
B_{k_1,k_2,k_3} &=& (D_+^{k_1} q) (D_+^{k_2} q)(D_+^{k_3} q); \quad 
  k_1+k_2+k_3 =N
\end{eqnarray} 
As well known, conformal symmetry allows one to resolve the mixing with 
operators containing total derivatives 
\cite{Brodsky:1980ny,Makeenko:1980bh,Ohrndorf:1981qv,Braun:1989iv,Mueller:1993hg,Belitsky:1998gc,Braun:2003rp}.
In particular, the coefficients $\varphi_{N,n}(\mu)$ with different values of $N$ (related to
the total conformal spin $J=N+3$) do not mix with each other by the one-loop evolution. 
The conformal symmetry is not sufficient, however, to solve the evolution equations:
the coefficients $\varphi_{N,n}(\mu)$ with the same $N$ but different $n$ do mix, 
producing a nontrivial spectrum of anomalous dimensions, see Fig.~1.
%
\begin{figure}[t]
\centerline{\resizebox{0.6\textwidth}{!}{
 \includegraphics{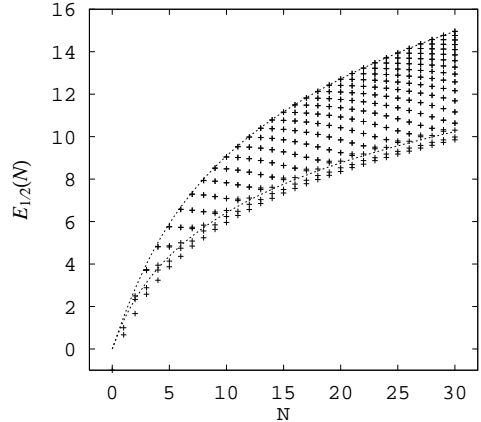}}}
\caption[]{\small The spectrum of anomalous dimensions  
$\gamma_N \equiv (1+1/N_c)E_N+3/2 C_F$ for the
baryon distribution amplitudes with helicity  $\lambda=1/2$. 
The lines of the largest and the smallest
eigenvalues for $\lambda=3/2$ are indicated by  dots
for comparison.}
\label{fig:e12}
\end{figure}
The corresponding multiplicatively renormalizeable contributions to the DA 
are given by linear combinations of the conformal polynomials
\begin{equation}
  P_{N,q}(x_i) = \sum_{n=0}^N c_{N,n}^{(q)} \Psi^{(12)3}_{N,n}(x_i)
\end{equation}
with the coefficients $c_{N,n}^{(q)}$ and anomalous dimensions $\gamma_{N,q}$ 
that have to be found by the diagonalization of the mixing matrix.

It turns out \cite{Braun:1998id} that the the index $q$ that enumerates the solutions
can be identified with an eigenvalue of a certain conserved
charge. The physical interpretation is that one is able to find a new 
`hidden' quantum number that distinguishes between partonic components 
in the proton with different scale dependence.

To explain this result, we have to introduce the so-called Hamiltonian approach \cite{Bukhvostov:1985rn}, 
in which the evolution kernels are rewritten in terms of the $SL(2)$ generators.
It is instructive to consider two cases separately, corresponding to helicity $\lambda=3/2$ and $\lambda=1/2$
operators related to the evolution of the $\Delta$-isobar DA and the nucleon, respectively.
The corresponding evolution kernels can be written in the following compact form  \cite{Braun:1998id,Braun:1999te}:
\begin{eqnarray}
 H_{3/2} &=& 2\left(1+\frac{1}{N_c}\right)\sum_{i<k}\Big[
 \psi(J_{ik})-\psi(2)\Big]+\frac32 C_F\,,
\label{3/2}\\
 H_{1/2} &=& H_{3/2}- 2\left(1+\frac{1}{N_c}\right)\left[
    \frac{1}{J_{12}(J_{12}-1)}+
\frac{1}{J_{23}(J_{23}-1)}\right].
\label{1/2}
\end{eqnarray}     
Here $\psi(x)$ is the logarithmic derivative of the $\Gamma$-function
and $J_{ik}, i,k =1,2,3$ are defined in terms of the two-particle
Casimir operators of the $SL(2,R)$ group   
\begin{eqnarray}
  J_{ik}(J_{ik}-1) &=& L_{ik}^2 \equiv (\vec L_i+\vec L_k)^2\,, 
\end{eqnarray}
with $\vec L_i$ being the group generators acting on  the 
i-th quark, which have to be taken in the adjoint representation~\cite{Braun:1999te}:
\begin{eqnarray}
 L_{k,0} P(x_i) &=& (x_k\partial_k + 1) P(x_i)\,,\nonumber\\
 L_{k,+} P(x_i) &=& - x_k P(x_i)\,,\nonumber\\
 L_{k,-} P(x_i) &=&  (x_k\partial_k^2 + 2\partial_k) P(x_i)\,.
\end{eqnarray}
Solution of the evolution equations corresponds in this language to 
solution of the Schr\"odinger equation
\begin{eqnarray}
  H P_{N,q}(x_i) &=& \gamma_{N,q} P_{N,q}(X_i)
\end{eqnarray} 
with $\gamma_{N,q}$ being the anomalous dimensions.
The $SL(2,R)$  invariance of the evolution equations implies 
that the generators of conformal transformations commute with 
the `Hamiltonians'
\begin{eqnarray}
 [H,L^2]&=& [H,L_\alpha] = 0\,,
\end{eqnarray} 
where $L^2 = (\vec L_1+\vec L_2+\vec L_3)^2$ and 
$L_\alpha = L_{1,\alpha}+L_{3,\alpha}+L_{3,\alpha}$, 
so that the polynomials $P_{N,q}(x_i)$
corresponding to multiplicatively renormalizable operators can be chosen 
simultaneously to be eigenfunctions of $L^2$ and $L_0$:
\begin{equation}
    L^2 P_{N,q} = (N+3)(N+2) P_{N,q}\,,\quad
    L_0 P_{N,q} = (N+3) P_{N,q}\,,\quad
    L_- P_{N,q} = 0\,.
\label{conform}   
\end{equation}
The third condition in (\ref{conform}) ensures that the operators do not 
contain overall total derivatives.

Main finding of Ref.~\refcite{Braun:1998id} is that the Hamiltonian $H_{3/2}$ possesses an 
additional integral of motion (conserved charge):
\begin{eqnarray}
  Q = \frac{i}{2}[L^2_{12},L^2_{23}] = i(\partial_1\!-\!\partial_2)
(\partial_2\!-\!\partial_3)(\partial_3\!-\!\partial_1) x_1 x_2 x_3
\,,
\quad [H_{3/2},Q] = 0\,. 
\end{eqnarray}
The evolution equation for baryon distribution functions with maximum 
helicity is, therefore, completely integrable. The premium is that 
instead of solving a Schr\"odinger equation with a complicated nonlocal 
Hamiltonian, it is sufficient to solve a much simpler equation
\begin{eqnarray}
    Q P_{N,q}(x_i) = q P_{N,q}(x_i)\,.
\label{Q}
\end{eqnarray}
Once the eigenfunctions are found, the eigenvalues of the Hamiltonian
(anomalous dimensions) are obtained as algebraic functions of $N,q$.

The Hamiltonian in (\ref{3/2}) is known as
the Hamiltonian of the so-called $XXX_{s=-1}$ Heisenberg spin magnet.  
The same Hamiltonian was encountered before in the interactions between 
reggeized gluons in QCD \cite{Lipatov:1994xy,Faddeev:1994zg}.

The equation in (\ref{Q}) cannot be solved exactly, but a wealth of analytic 
results can be obtained by means of the $1/N$ expansion \cite{Korchemsky}. 
One general consequence of complete integrability is that all anomalous dimensions are double degenerate
except for the lowest ones for each {\it even} $N$, corresponding to 
the solution with $q=0$. The corresponding eigenfunctions have a very 
simple form \cite{Braun:1999te}
\begin{eqnarray}
x_1 x_2 x_3 P^{\lambda=3/2}_{N,q=0}(x_i) &=& x_1(1-x_1)C^{3/2}_{N+1}(1-2x_1)
+x_2(1-x_2)C^{3/2}_{N+1}(1-2x_2)
\nonumber\\&&{}+x_3(1-x_3)C^{3/2}_{N+1}(1-2x_3)
\label{lowest}
\end{eqnarray}
and the  anomalous dimension is equal to 
\begin{eqnarray}
 \gamma_{N,q=0} &=& \left(1+1/N_c\right)
   \Big[4\psi(N+3)+4\gamma_E-6\Big]
  +3/2 C_F\,.
\label{lowen}
\end{eqnarray}
The asymptotic expansions    
for the charge $q$ and the anomalous dimensions at large $N$ are available 
to the order $1/N^8$ \cite{Korchemsky,Braun:1999te} and give very accurate results. 

The additional term in $H_{1/2}$ (the nucleon) spoils integrability but can be considered
as a (calculable) small correction for all of the spectrum except for 
two lowest levels \cite{Braun:1999te}. To illustrate this,
consider the flow of energy levels for the Hamiltonian 
 $H(\epsilon) = \sum_{i<k}\Big[
 \psi(J_{ik})-\psi(2)\Big] - \epsilon \Big[1/L_{12}^2+1/L_{23}^2\Big]$
(cf. (\ref{3/2}), (\ref{1/2}))
as a function of an auxiliary parameter $\epsilon$, see Fig.~2.
\begin{figure}[t]
\centerline{\resizebox{0.6\textwidth}{!}{
 \includegraphics{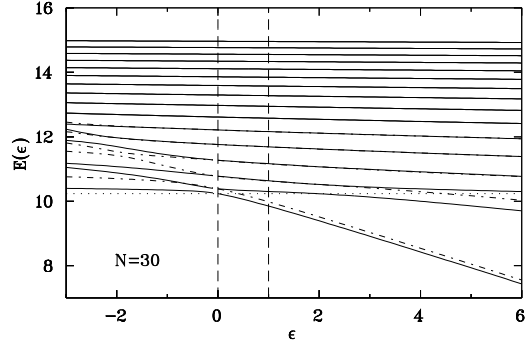}}}
\caption[]{\small The flow of energy eigenvalues for the Hamiltonian
   ${H}(\epsilon)$ for $N=30$.
The solid and the dash-dotted curves show
the parity-even and parity-odd levels, respectively.
The two vertical dashed lines
indicate ${H}_{3/2}\equiv{ H}(\epsilon=0)$ and
${H}_{1/2}\equiv{H}(\epsilon=1)$, respectively (up to the color factors). 
The horizontal dotted line shows position of the unperturbed `ground state' given by
Eq.~(\protect{\ref{lowen}}).
}
\label{figure2}
\end{figure}
It is seen that the two lowest levels decouple from the rest of the 
spectrum and are separated from it by a finite mass gap. As shown in 
Ref.~\refcite{Braun:1999te}, this phenomenon can be interpreted as binding of the 
two quarks with opposite helicity and forming a scalar ``diquark''.
The effective Hamiltonian for the low-lying levels can be constructed 
and turns out to be a generalization of the famous Kroning-Penney problem
for a particle in a $\delta$-function type periodic potential.
The value of the mass gap between the lowest and the next-to-lowest 
anomalous dimensions at $N\to\infty$ can be calculated combining the small-$\epsilon$
and the large-$\epsilon$ expansions and is equal to 
 $ \Delta\gamma = 0.32\cdot(1+1/N_c) $
in agreement with the direct numerical calculations.
The corresponding contributions to the nucleon DA are of the 
form \cite{Braun:1999te}
\begin{eqnarray}
 P^{\lambda=1/2}_{N,q=0}(x_i) &\stackrel{\ln N\to \infty}{=}& P^{(1,3)}_N(1-2x_3)\pm  P^{(1,3)}_N(1-2x_1)\,, 
\end{eqnarray}
where $P^{(1,3)}_N(x)$ are Jacobi polynomials.

The approach based on complete integrability can be used to obtain parts of the two-loop
evolution kernels for baryon operators beyond the leading order \cite{Belitsky:2004sf},
but a complete calculation to the two-loop accuracy is so far absent.

\section{Nonperturbative parameters}

To the leading-order accuracy in the conformal spin expansion, the leading-twist-3
DA involves one, $f_N$,  and the twist-4 DAs two, $\lambda_1$ and $\lambda_2$, nonperturbative parameters,
To the next-to-leading accuracy in the conformal spin there are two additional parameters for twist-3, 
$A_1^u$ and $V_1^d$, and three parameters for twist-4, $f_1^u$, $f_1^d$ and $f_2^d $, cf. Eqs.~(\ref{par3}),(\ref{par4}). 
The number of parameters proliferates rapidly if higher spins are included, and their estimates 
become increasingly complicated and unreliable. Hence I stop at the first nontrivial order and
summarize the existing estimates in Table~1 and Table~2 for the leading and the higher twist DAs,
respectively.

Most of the estimates are obtained using QCD sum rules. 
The quoted numbers correspond to the sum rules to the leading order accuracy
in the QCD coupling. The NLO radiative corrections are known for $\lambda_1$  
\cite{Jamin:1987gq,Ovchinnikov:1988zx} but not for other cases, to my knowledge.
The effect of such corrections can be substantial, see e.g. Ref.~\refcite{Sadovnikova:2005ye}.

\begin{table}[t]
\tbl{Parameters of the leading-twist nucleon distribution amplitude (\ref{DA3}), (\ref{par3})  
 at the scale 1 GeV. The constant $f_N$ is  given in units of \protect{$10^{-3}$~GeV$^2$}}
{\begin{tabular}{l|c|l|l|l}\hline
Method & $f_N$               & $V_1^d            $      & $A_1^u$    & Ref.    \\ \hline
asymptotic  & $-$            &  $1/3$    & $ 0$    & \\ 
QCDSR  & $5.3\pm 0.5$  &  $0.220$  & $0.480$ &[\refcite{Chernyak:1983ej}]  \\ 
QCDSR  & $5.0\pm 0.3$  &  $0.229$  & $0.387$ &[\refcite{Chernyak:1987nv}] \\ 
QCDSR  & $5.1\pm 0.3$  &  $ 0.240$  & $0.340$ &[\refcite{King:1986wi}]  \\ 
QCDSR  & $-$  &  $0.236$  & $0.490$  &[\refcite{Gari:1986dr}]  \\ 
Model  & $-$                 &  $0.310$  & $0.071$  &[\refcite{Bolz:1996sw}] \\ 
LCSR   & $-$                 &  $0.300$  & $0.130$  &[\refcite{Braun:2006hz}] \\ 
\hline  
\end{tabular}}
\end{table}

\begin{table}[t]
\tbl{Parameters of the twist-four nucleon distribution amplitudes (\ref{DA4}), (\ref{par4})  
 at the scale 1 GeV. The constants $\lambda_1$ and $\lambda_2$ are given in units of 
\protect{$10^{-3}$~GeV$^2$}}
{\begin{tabular}{l|c|l|l|l|l|l}\hline
Method & $\lambda_1$ & $\lambda_2$ & $f_1^d$   & $f_2^d$  & $f_1^u$    & Ref.    \\ \hline
asymptotic  & $-$             & $-$                 &    3/10    & 4/15 & 1/10 & \\ 
QCDSR       & $-27\pm 5$  &  $ 54\pm 19 $  & $0.40\pm 0.05$  
& $0.22\pm0.05$ & $0.07\pm 0.05$ &[\refcite{Braun:2006hz}] \\ 
LCSR   & $-$ & $-$  & $0.33$  
& $0.25$ & $0.09$  &[\refcite{Braun:2006hz}] \\ 
\hline  
\end{tabular}}
\end{table}

The calculations presented 
in Refs.~\refcite{Chernyak:1983ej,Chernyak:1987nv,King:1986wi,Gari:1986dr}
make use of the same sum rule and are, therefore, not entirely independent.
The errors are difficult to quantify, but are probably of the size of 
the spread in the quoted values. 
The result for $f_N$ is expected to be rather reliable although the quoted 
error might well be underestimated.    
The parameter $\lambda_1$ is also well known to QCD sum rule
practitioners and corresponds to the nucleon coupling to the so-called 
Ioffe current \cite{Ioffe:1981kw}. Note that although an overall sign in the 
couplings $f_N, \lambda_{1,2}$ is arbitrary and can be readjusted by the 
phase factor in the nucleon wave function, the relative sign is physical and important
for the applications. 

Alternatively, there exists a phenomenological model for the leading-twist DA \cite{Bolz:1996sw}
which was obtained by modelling the soft contribution to electromagnetic form factors by a 
convolution of light-cone wave functions. Estimates of the higher-twist DAs in the same 
technique are not available.  Finally, I quote the parameters obtained in Ref.~\refcite{Braun:2006hz}
from the fit of the light-cone sum rules to the experimental data on the nucleon form factors.
This approach will be explained below.  In future, one should expect that 
a few lowest order parameters in the conformal expansion of baryon DAs can be calculated on the lattice, 
cf. Ref.~\refcite{Martinelli:1988xs}. Main technical problem 
on this way seems to be the necessity to use nonperturbative renormalization of three-quark operators.

\section{Light-Cone Sum Rules}

Main problem that does not allow to extract the information on baryon DAs from experiment is that 
nature does not provide us with point-like three-quark currents. 
Baryon number conservation implies that physical processes always involve baryons in pairs. Hence 
one has to deal with the convolution of two baryon DAs, and also the so-called ``soft'' contributions to the
form factors which cannot expressed in terms of DAs prove to be numerically significant at present energies. 
A (partial) remedy is suggested by the approach known as light-cone 
sum rules (LCSRs) \cite{Balitsky:1989ry,Braun:1988qv,Chernyak:1990ag}.
This technique is attractive because in LCSRs  ``soft'' contributions to the form factors are calculated in 
terms of the same DAs that
enter the pQCD calculation and there is no double counting. Thus, the LCSRs provide one with the most 
direct relation of the hadron form factors
and distribution amplitudes that is available at present, with no other nonperturbative parameters.  
 
\begin{figure}[t]
\centerline{\epsfxsize3.5cm\epsffile{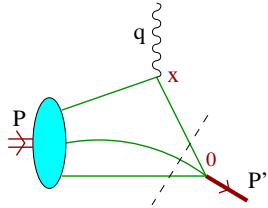}}
\caption{\label{figsum}\small
Schematic structure of the light-cone sum rule for baryon form factors.
}
\end{figure}

The basic object of the LCSR approach is the 
correlation function 
\[\int\! dx\, e^{-iqx}\langle 0| T \{ \eta (0) j(x) \} | N(P) \rangle \] in which
$j$  represents the electromagnetic (or weak) probe and $\eta$
is a suitable operator with nucleon quantum numbers.    
 The other (in this example, initial state) nucleon  is explicitly represented by its state vector 
 $| N(P)\rangle $, see a schematic representation in Fig.~\ref{figsum}.
 When both  the momentum transfer  $Q^2$ and 
 the momentum $(P')^2 = (P-q)^2$ flowing in the $\eta$ vertex are large and negative,
 the asymptotics of the correlation function is governed by the light-cone kinematics $x^2\to 0$ and
 can be studied using the operator product expansion (OPE)   
$T \{ \eta(0) j(x) \} \sim \sum C_i(x) {\cal O}_i(0)$ on the 
light-cone $x^2=0$.   The  $x^2$-singularity  of a particular perturbatively calculable
short-distance factor  $C_i(x)$  is determined by the twist of the relevant
composite operator ${\cal O}_i$, whose matrix element $\langle 0|  {\cal O}_i(0)| N(P) \rangle $
is given by an appropriate moment of the nucleon DA.
Next, one can represent the answer in form of the dispersion integral in $(P')^2$ and define the nucleon contribution
by the cutoff in the quark-antiquark invariant mass, the so-called interval of duality $s_0$ (or continuum threshold).
The main role of the interval of duality is that it does not allow large momenta $|k^2| > s_0$ to flow through the 
 $\eta$-vertex; to the lowest order $O(\alpha_s^0)$ one obtains a purely soft    
contribution to the form factor as a sum of terms ordered by twist of the relevant operators and
hence including both the leading- and the higher-twist nucleon DAs. Note that, in difference to the hard mechanism, the 
contribution of higher-twist DAs is only suppressed by powers of 
$|(P')^2|\sim 1-2$~GeV$^2$ (which is translated to the suppression
by powers of the Borel parameter after applying the usual QCD sum rule machinery), but not by powers of $Q^2$. This feature is
in agreement with the common wisdom that soft contributions are not constrained to small transverse separations.  

The LCSR expansion also  contains terms  
generating the asymptotic pQCD contributions. They   appear 
at proper order in $\alpha_s$, i.e., in  the $O(\alpha_s)$ term for the
pion form factor, at the $\alpha_s^2$ order for the nucleon form factors, etc. 
In the pion case, it was explicitly demonstrated 
\cite{Braun:1999uj,Bijnens:2002mg} that the contribution of hard  
rescattering is correctly reproduced in the LCSR   
approach as a part of the $O(\alpha_s)$ correction.
It should be noted that  the  diagrams of LCSR that 
contain the ``hard'' pQCD  contributions also possess ``soft'' parts,
i.e., one should perform  a separation  of ``hard'' and ``soft''
terms inside each diagram.  As a result, 
the distinction between ``hard'' and ``soft'' contributions appears to 
be scale- and scheme-dependent \cite{Braun:1999uj}. 
During the  last years there have been numerous applications of LCSRs  
to mesons, see Refs.~\refcite{Braun:1997kw,Colangelo:2000dp} for a review.
{}Following the work Ref.~\refcite{Braun:2001tj} nucleon form factors 
were further considered in this framework in Refs.~\refcite{Wang:2006uv,Wang:2006su,Lenz:2003tq,Braun:2006hz} 
and the weak decays $\Lambda_b\to p\ell\nu_\ell$, $\Lambda_c\to \Lambda\ell\nu_\ell$ in
Refs.~\refcite{Huang:2004vf,Huang:2006ny}.  The generalization to the $N\gamma\Delta$ transition form factor
was worked out in Ref.~\refcite{Braun:2005be}.

\begin{figure}[t]
\begin{center}
  \includegraphics[width=0.44\textwidth,angle=0]{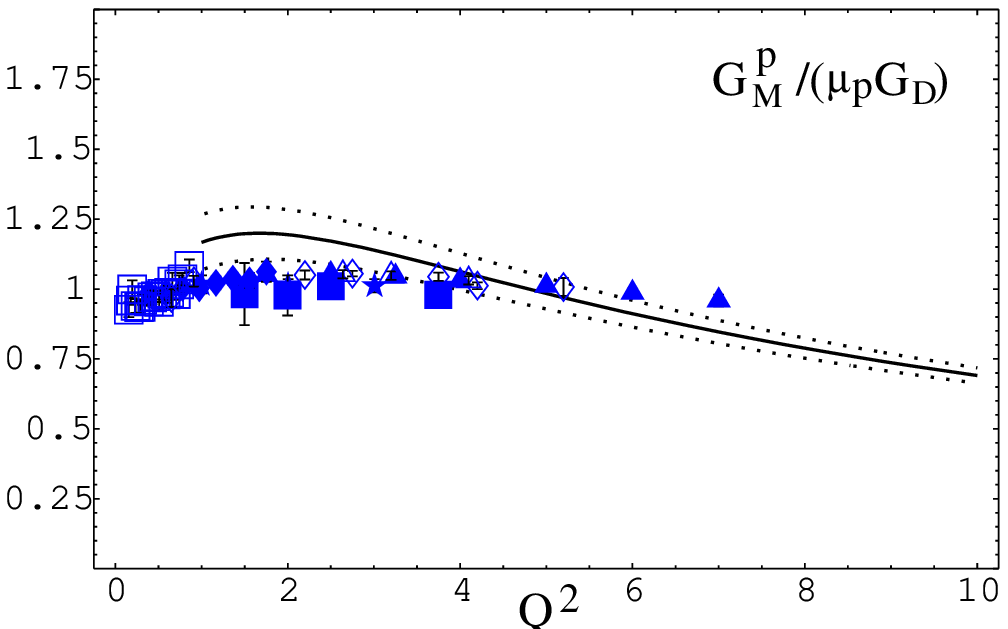}
  \includegraphics[width=0.44\textwidth,angle=0]{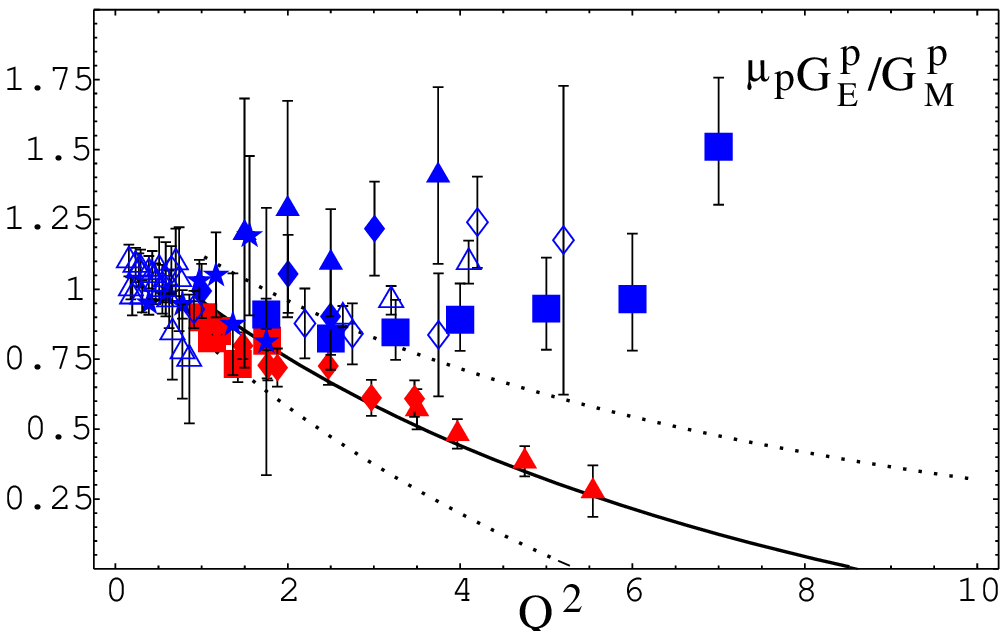}\\[1mm]
  \includegraphics[width=0.44\textwidth,angle=0]{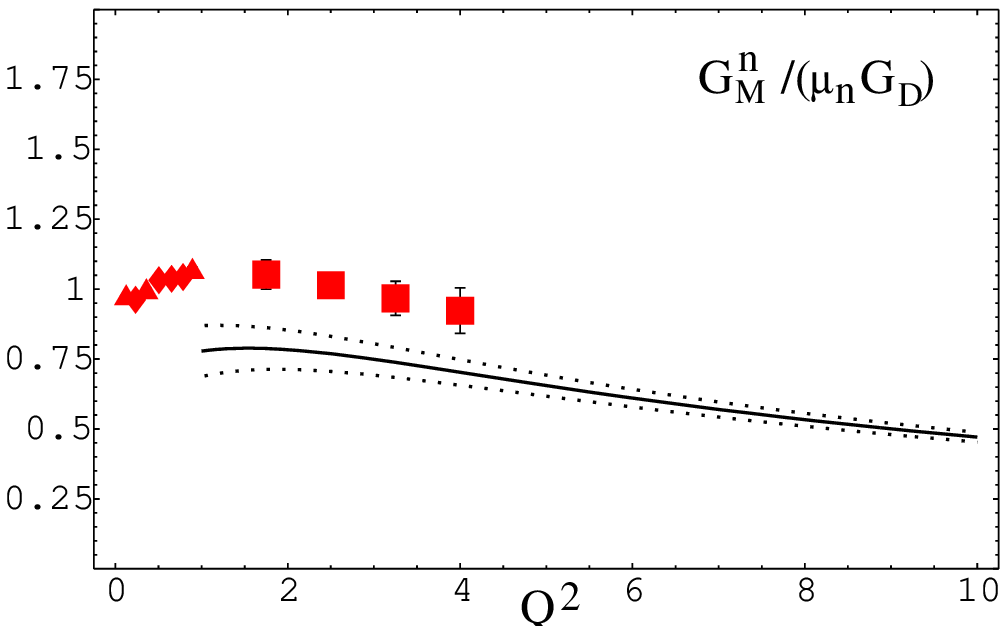}
  \includegraphics[width=0.44\textwidth,angle=0]{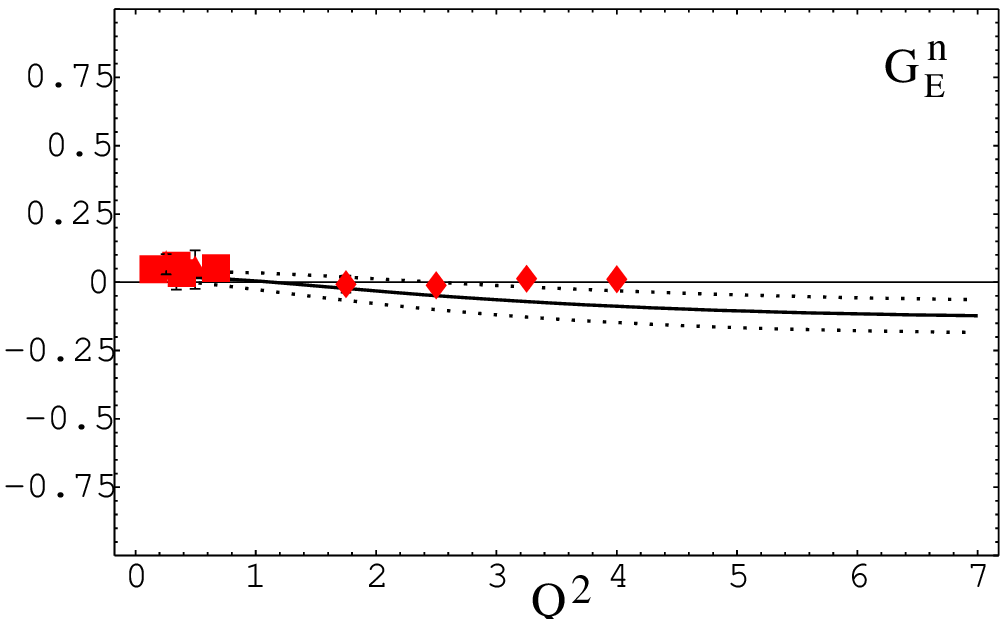}
\end{center}
\caption{LCSR results (solid curves) for the electromagnetic form factors of the 
nucleon, obtained using the model of the nucleon DAs with parameters from Tables~1,2.
The dotted curves show the effect of the variation of the ratio $f_N/\lambda_1$ by 30\%.
For the identification of the data points and details of the calculation see
Ref.~36.
}
\label{fig:Femtune}
\end{figure}

The net outcome of these studies is that all nucleon form factors (with an exception of the 
magnetic $N\Delta\gamma$ transition) can be reproduced to roughly 20\% accuracy by using 
the parameters of the proton DA summarized in Tables 1,2 above, which are roughly in the 
middle of the range between asymptotic DAs and the QCD sum rule predictions, see
Fig.~\ref{fig:Femtune}.  This conclusion is preliminary, however. 
More studies are needed and in particular radiative 
corrections to the sum rules have to be calculated.    

\section{Conclusions}

Baryon distribution amplitudes are fundamental nonperturbative 
functions describing the hadron structure in configurations
with a fixed number of Fock constituents at small transverse 
separation. They are equally important and to a large extent complementary to conventional 
parton distributions which correspond to one-particle probability
distributions for the parton momentum fraction in an average  configuration.
The theory of baryon DAs has reached a certain degree of maturity. In particular,
their scale dependence is well understood and reveals a beautiful hidden symmetry
of QCD which is not seen at the level of the QCD Lagrangian. The basic tool 
to describe DAs is provided by the conformal expansion combined with EOM (for higher
twists) that allows one to obtain parameterizations with the minimum number 
of nonperturbative parameters. There are indications that the conformal expansion
is converging sufficiently rapidly so that only a few terms are needed for most 
of the practical purposes. A qualitative picture inspired by the QCD sum rule 
calculations \cite{Chernyak:1983ej} seems to be that the valence quark with the spin parallel to
that of the proton carries most of its momentum. It is timely to make this picture quantitative;
combination of LCSRs and lattice calculations should allow one to determine momentum fractions carried 
by the three valence quarks with 5-7\% precision within a few years.
Further progress will depend decisively on whether studies of hard exclusive processes can be made fully 
quantitative. High quality data are needed in the $Q^2\sim 10$~GeV$^2$ range, and 
one has to develop a consistent theoretical framework for the treatment of end-point contributions.
  
\section*{Acknowledgements}

The author is grateful to I.~Balitsky, P.~Ball, S.~Derkachov, R.~Fries, 
G.~Korchemsky, A.~Lenz, A.~Manashov, N.~Mahnke, D.~Mueller, A.~Peters, G.~Peters,
A.~Radyushkin, E.~Stein and M.~Wittmann  for the collaboration on the subject of this review.
Special thanks are  to the organizers of CAQCD06 for the invitation and hospitality.

\end{document}